\documentclass[%
 aip,
 amsmath,amssymb,
 reprint,%
]{revtex4-1}

\usepackage{graphicx}
\usepackage{dcolumn}
\usepackage{bm}

\usepackage[utf8]{inputenc}
\usepackage[T1]{fontenc}
\usepackage{mathptmx}
\usepackage{etoolbox}

\usepackage{xcolor}
\usepackage{soul}
\usepackage[colorlinks=true,linkcolor=blue,citecolor=blue,urlcolor=blue]{hyperref}

\newcommand*{\figreff}[2][]{%
  \hyperref[{fig:#2}]{%
    Figure~\ref*{fig:#2}%
    \ifx\\#1\\%
    \else
      \,#1%
    \fi
  }%
}

\newcommand*{\figref}[2][]{%
  \hyperref[{fig:#2}]{%
    Fig.~\ref*{fig:#2}%
    \ifx\\#1\\%
    \else
      \,#1%
    \fi
  }%
}

\makeatletter
    \def\@email#1#2{%
 \endgroup
 \patchcmd{\titleblock@produce}
  {\frontmatter@RRAPformat}
  {\frontmatter@RRAPformat{\produce@RRAP{*#1\href{mailto:#2}{#2}}}\frontmatter@RRAPformat}
  {}{}
}%
\makeatother
\begin{document}

\preprint{AIP/123-QED}

\title{Phase Field Simulation of Liquid Filling on Grooved Surfaces for \\Complete, Partial and Pseudo-partial Wetting Cases}
\author{Fandi Oktasendra}
 \affiliation{Physics Department, Durham University, Durham DH1 3LE, United Kingdom}
\affiliation{Physics Department, Universitas Negeri Padang, Padang 25131, Indonesia }

\author{Arben Jusufi}%
\affiliation{Research, ExxonMobil Technology and Engineering Company, Annandale, NJ 08801, USA}

\author{Andrew R. Konicek}%
\affiliation{Research, ExxonMobil Technology and Engineering Company, Annandale, NJ 08801, USA}

\author{Mohsen S. Yeganeh}%
\affiliation{Research, ExxonMobil Technology and Engineering Company, Annandale, NJ 08801, USA}

\author{Jack R. Panter}%
\affiliation{School of Engineering, University of East Anglia, Norwich NR4 7TJ, United Kingdom}

\author{Halim Kusumaatmaja}
 \affiliation{Physics Department, Durham University, Durham DH1 3LE, United Kingdom}

 \email[Authors to whom correspondence should be addressed: ]{halim.kusumaatmaja@durham.ac.uk and Jack.Panter@uea.ac.uk}

\date{\today}

\begin{abstract}
We develop and harness a phase field simulation method to study liquid filling on grooved surfaces. We consider both short-range and long-range liquid-solid interactions, with the latter including purely attractive and repulsive interactions, as well as those with short-range attraction and long-range repulsion. This allows us to capture complete, partial and pseudo-partial wetting states, demonstrating complex disjoining pressure profiles over the full range of possible contact angles as previously proposed in the literature. Applying the simulation method to study liquid filling on grooved surfaces, we compare the filling transition for the three different classes of wetting states as we vary the pressure difference between the liquid and gas phases. The filling and emptying transition is reversible for the complete wetting case, while significant hysteresis is observed for the partial and pseudo-partial cases. In agreement with previous studies, we also show that the critical pressure for the filling transition follows the Kelvin equation for the complete and partial wetting scenarios. Finally, we find the filling transition can display a number of distinct morphological pathways for the pseudo-partial wetting cases, as we demonstrate here for varying groove dimensions.

\end{abstract}

\maketitle

%

\section{\label{sec:level1}Introduction \protect} 

Wetting of solid surfaces by liquids is ubiquitous in nature and critically important for many technological and industrial applications ranging from printing, coating, microfluidics to oil recovery and carbon capture~\cite{Bonn2009,Wen2015,Wang2016,mohsen2022}. Given the importance of surface wettability, along with rapid advances in surface engineering techniques, such as lithography, 3D printing, and surface self-assembly~\cite{Yan2020,Brassat2020,Kong2019}, understanding the roles of surface topography on the wetting behavior of liquids has emerged as a prominent area of research. 

Numerous works to date have investigated how surface structures can give rise to advantageous surface wettability~\cite{Liu2017,Zhu2021,Lv2021,Song2023}, including superhydrophobicity, self-cleaning, drag reduction, and directional spreading. However, the majority of these studies take a macroscopic view of wetting phenomena where the liquid-solid-gas interactions are represented by a single parameter describing the contact angle. At the same time, it is well established in the literature that the intermolecular interactions between the liquid and solid molecules can be highly complex~\cite{Bonn2009,israelachvili2011intermolecular,Rauscher2008}.

Such intermolecular interactions include hydrogen bonds, van der Waals, dipole-dipole interactions, and others~\cite{Brochard-Wyart1991}. However, their effect on wetting can be understood by looking at a thin liquid film of thickness \textit{e} on a solid substrate, from which all intermolecular interactions can be incorporated in terms of an effective interface potential. This is defined as the cost of energy per unit area to maintain the thin film at a given thickness~\cite{Rauscher2008}. From this effective interface potential, one can derive the effective repulsive force per unit area between the solid-liquid and liquid-gas interfaces, known as the disjoining pressure~\cite{Bonn2009}.

When considering phenomena at length scales smaller or comparable to the range of the effective interface potential (typically, of the order of hundreds of nanometres~\cite{Rauscher2008,Grzelak2010}), a contact angle description of the three-phase interaction in itself is not adequate. It is important to consider the distance-dependent interactions from the solid surface. Indeed, depending on the disjoining pressure profiles (which capture the aforementioned distance-dependent interactions), different wetting states can arise~\cite{Brochard-Wyart1991,silberzan1991evidence}: complete wetting, where liquid fully spreads on the solid surface; partial wetting, where a finite contact angle is formed at the three-phase solid-liquid-gas contact line; and pseudo-partial wetting, where the macroscopic liquid domain (\textit{i.e.} droplet) is surrounded by a thin liquid film.

Previous approaches to computationally study nanoscale fluid phenomena incorporate atomistic details, such as using Molecular Dynamics (MD)~\cite{Sethi2022} and Density Functional Theory (DFT)~\cite{Malijevsky2018,Giacomello2016,Singh2022,Hughes2015,Malijevsky2013}. Here, we show how a mesoscale model, the phase field model, can be augmented to enable a wide variety of short and long-range solid-fluid interactions described above. This is distinct from previous phase field wetting simulations, which typically treat wetting as a boundary condition at a solid surface \cite{cahn1977critical,Jacqmin1999}, and neglect long-range forces. Since we are interested in static and quasi-static phenomena in this paper, we will directly minimize the free energy of the phase field models.

As phase-field models are computationally less demanding than traditional nanoscale methods, the incorporation of long-range interactions should allow highly complicated structures to be studied. This is relevant not only because smaller and more complex features can be reliably manufactured~\cite{MacGregor-Ramiasa2017}; but also because they are key for the emergence of interfacial phase transitions, such as liquid adsorption and liquid filling~\cite{Singh2022,Bormashenko2013,derjaguin1992theory,Giacomello2016b}, which start at the smallest surface features. Such phase transitions are important for many applications, such as thin film condensation and evaporation~\cite{enright2013condensation,lu2021surface}, and heat transfer~\cite{Cao2020}.

To demonstrate the versatility of the phase field method, we apply it to study liquid filling and emptying on grooved surfaces as the liquid pressure is varied~\cite{Malijevsky2012,Parry2014,Malijevsky2018}. We will compare the results for short-range and long-range liquid-solid interactions. We will also contrast them for complete, partial, and pseudo-partial wetting scenarios. To the best of our knowledge, this is the first systematic liquid filling transition study for the pseudo-partial wetting case. Due to the competition between short-range attraction and long-range repulsion, it leads to several possible pathways and critical pressure dependence on geometry that are distinct from the complete and partial wetting cases.

 This paper is structured as follows. In Sec.~\ref{sec.2}, the theoretical basis of the model is introduced. We present our main results in Sec.~\ref{sec.3}, which contains two sub-sections. The first part of Sec.~\ref{sec.3} is devoted to simulation results on a flat surface, while the second part is for grooved surfaces. We summarize our work and discuss avenues for future work in Sec.~\ref{sec.4}.  

\section{\label{sec.2}Phase Field Method}
We use a phase field model to describe a binary fluid system in contact with solid surfaces. In this model, the scalar order parameter $\phi(\textbf{r})$ is used to represent the local composition of the fluid with $\phi = 1$ and $ -1 $ indicating the liquid phase and pure gas phase, respectively. The equilibrium phase profile is obtained by minimizing the total free energy \cite{Kusumaatmaja2015,Panter2017}
\begin{equation}
\Psi = \Psi_\textrm{b} + \Psi_{\textrm{s}} + \Psi_{\textrm{c}}. 
\label{eq:total_free_energy}
\end{equation}
Here, $\Psi_\textrm{b}$, $\Psi_\textrm{s}$, and $\Psi_\textrm{c}$ are the liquid bulk, liquid-solid surface, and pressure or volume constraint terms. More specifically, $\Psi_\textrm{b}$ is the free energy contribution arising from a binary fluid system describing the homogeneous (bulk) and liquid-gas interface,
\begin{equation}
\Psi_\textrm{b} = \int_V {\left [ {
		{\frac{1}{4 \epsilon}\left( {\phi ^2  - 1 } \right)^2} + \frac{\epsilon}{2}\left| {\nabla \phi } \right|^2 } \right ]\textrm{d}V},
\label{eq:free_energy_bulk}
\end{equation}
where $\epsilon$ is the width of the liquid-gas interface. In this model, the liquid-gas surface tension takes the value of 
\begin{equation}
\gamma_{\textrm{lg}} = \sqrt{8/9}.
\label{eq:surface_tension}
\end{equation}

The surface energy contribution in the total free energy, $\Psi_{\textrm{s}}$, comes from interactions between liquid and solid, which are responsible for determining the wettability of the liquid on solid surfaces. Here, we explore describing the liquid-solid interactions in two ways. First, we employ long-range solid-liquid interactions to mimic the complexity of intermolecular interactions between liquid and solid. The surface energy can be written as
\begin{equation}
\Psi_{\textrm{s}} = \int_{V} {{\cal F}_{\textrm{s}}(\textbf{r}) f(\phi)  \textrm{d}V }.
\label{eq:free_energy_surface_LR}
\end{equation} 
${\cal F}_{\textrm{s}}(\textbf{r})$ is the energy density due to long-range interactions between liquid and solid separated by a distance \textbf{r}. We are free to choose any ${\cal F}_{\textrm{s}}(\textbf{r})$ we desire, but for this work we choose
\begin{equation}
{\cal F}_{\textrm{s}}(\textbf{r})= \int_{V_\textrm{s}} {u_{\textrm{lr}}(\textbf{r})\textrm{d}V_\textrm{s}},
\label{eq:lr-external}
\end{equation} 
where $u_{\textrm{lr}}(\textbf{r})$, namely the effective interaction, is integrated over the volume of solid, $V_s$. The effective interactions $u_{\textrm{lr}}(\textbf{r})$ can take one of the following forms
\begin{equation}
u_{\textrm{lr}}^{1}\left( \textbf{r} \right) =  \frac{\alpha}{{ \left( \beta + \left| {\textbf{r}-\textbf{r}_\textrm{s}} \right| \right) ^6 }},  
\label{eq:lr-1}
\end{equation}
or
\begin{equation}
u_{\textrm{lr}}^{2}(\mathbf{r}) = \left\{\begin{matrix}
-u_\textrm{o}  & \quad ,|\mathbf{r}-\mathbf{r}_\textrm{s}| < \sigma \quad \\
u_\textrm{p} \left( \frac{\sigma}{|\mathbf{r} -\mathbf{r}_\textrm{s}|} \right)^6  & \quad ,|\mathbf{r}-\mathbf{r}_\textrm{s}| \geq  \sigma, \quad
\end{matrix}\right.
\label{eq:lr-2}
\end{equation}
in which $ \alpha$, $\beta $, $\sigma$, $u_\textrm{o}$ and $u_\textrm{p}$ are parameters in the models and $\textbf{r}_\textrm{s}$ is the coordinate position in the solid. The effective interaction $u_{\textrm{lr}}^{1}$ can be either attractive or repulsive depending on the sign of $\alpha$. The parameter $\beta$, which is taken to be positive, is used to avoid $u_{\textrm{lr}}^{1}$ going to infinity for $ \left| \textbf{r} - \textbf{r}_\textrm{s} \right| = 0 $ and to control the width of the decaying interaction $u_{\textrm{lr}}^{1}$. The form of $u_{\textrm{lr}}^{2}$ is designed to have an attractive interaction near the surface with a finite value of $-u_\textrm{o}$ and a repulsive interaction far from the surface with a maximum value of $u_\textrm{p}$ at $ |\mathbf{r} -\mathbf{r}_\textrm{s}| = \sigma $. The effective interaction $u_{\textrm{lr}}^{2}$ is chosen so that, when integrated, the long-range liquid-solid interaction is repulsive close to the solid surface and attractive far from the surface.

To ensure this interaction energy density is only contributed by the interaction between liquid and solid, it must be modulated by the local fluid composition, such that the liquid phase should experience the full ${\cal F}_{\textrm{s}}(\textbf{r})$, and the gas phase should not experience ${\cal F}_{\textrm{s}}(\textbf{r})$. For this purpose, we use $f(\phi)$, which is a fourth order polynomial that switches ${\cal F}_{\textrm{s}}(\textbf{r})$ between the liquid and gas phases, given by
\begin{equation}
{f(\phi)=  {\frac{3}{4}\phi  - \frac{1}{3}t(\textbf{r})\phi ^2  - \frac{1}{4}\phi ^3  + \frac{1}{6}t(\textbf{r})\phi ^4 + \frac{1}{6}(t(\textbf{r})+3)} },
\label{eq:lr-quartic}
\end{equation} 
with $t(\textbf{r})=-{\mathrm{sign}}({\cal F}_{\textrm{s}}(\textbf{r}))$. We choose the form in Eq.~(\ref{eq:lr-quartic}) because it prevents the enrichment of one of the phases at the surface owing to the following features: (i) $\frac{\textrm{d}\Psi_{\textrm{s}}}{\textrm{d}\phi} = 0$ at the bulk equilibrium values of $\phi = \pm 1$, (ii) $\Psi_{\textrm{s}}$ increases monotonically with $|\phi|$ for $|\phi| > 1$ which gives an energy penalty to the total free energy, and (iii) $\Psi_{\textrm{s}}$ is globally minimized at $\phi = +1$ (\textit{i.e.} the liquid phase) for ${\cal F}_{\textrm{s}}(\textbf{r}) < 0$ and at $\phi = -1$ (\textit{i.e.} the gas phase) for ${\cal F}_{\textrm{s}}(\textbf{r}) > 0$. Without the enrichment at the surface, we are able to maintain the simulation stability as well as approximate the fluid incompressibility.

The second way to introduce the liquid-solid interaction, following Cahn \cite{cahn1977critical}, is to use a short-range interaction between liquid and solid at the surface, which can be approximated by an integral over the solid surface area \textit{A},
\begin{equation}
\Psi_{\textrm{s}} = \int_{A} {\cal F}_{\textrm{s}}(\textbf{r}_\textrm{s}) f(\phi_\textrm{s}) \textrm{d}A, 
\label{eq:free_energy_surface_SR}
\end{equation} 
where we recall that $f(\phi_{\textrm{s}})$ is the polynomial form in Eq.~(\ref{eq:lr-quartic}) for $\phi = \phi_\textrm{s}$, where $\phi_\textrm{s}$ is the value of $\phi$ at the solid surface. To calculate the liquid-solid energy density at the surface, $ {\cal F}_{\textrm{s}}(\textbf{r}_\textrm{s}) $, we relate its value to the gas-solid, liquid-solid, and liquid-gas surface tensions, $\gamma_{\textrm{gs}}, \gamma_{\textrm{ls}}$ and $\gamma_{\textrm{lg}}$ respectively, via the spreading parameter \textit{S}, where $S = \gamma_{\textrm{gs}}-\gamma_{\textrm{ls}}-\gamma_{\textrm{lg}}$. The surface energy density must be equal to the gas-solid surface tension when the surface is completely dry  $( {\cal F}_{\textrm{s}}(\textbf{r}_\textrm{s})f(\phi_\textrm{s}=-1) = \gamma_{\textrm{gs}})$, and be equal to the liquid-solid surface tension when the surface is completely wet $({\cal F}_{\textrm{s}}(\textbf{r}_\textrm{s})f(\phi_\textrm{s}=+1)=\gamma_{\textrm{ls}})$. Therefore, from Eq.~(\ref{eq:lr-quartic}) we can have the relation $\gamma_{\textrm{gs}} - \gamma_{\textrm{ls}} = -{\cal F}_{\textrm{s}}(\textbf{r}_\textrm{s})$, independent of the value of $t(\textbf{r})$ leading to 
\begin{equation}
{\cal F}_{\textrm{s}}(\textbf{r}_\textrm{s}) = - S - \gamma_{\textrm{lg}}.
\label{eq:free_energy_surface_SR_1}
\end{equation}
For partial wetting ($S < 0$), we can also relate ${\cal F}_{\textrm{s}}(\textbf{r}_\textrm{s})$ to the contact angle $\theta$ via Young equation, $\gamma_{\textrm{lg}}\cos \theta = \gamma_{\textrm{gs}}-\gamma_{\textrm{ls}}$, yielding 
\begin{equation}
{\cal F}_{\textrm{s}}(\textbf{r}_\textrm{s}) = - \sqrt{8/9} \cos \theta,
\label{eq:free_energy_surface_SR_2}
\end{equation}
where we have substituted $\gamma_{\textrm{lg}} = \sqrt{8/9}$ from Eq.~(\ref{eq:surface_tension}).

The last term in the free energy in Eq.~(\ref{eq:total_free_energy}), $\Psi_\textrm{c}$, reflects the constraint applied to the system, which can be chosen to either define the pressure difference between the liquid and gas, or constrain the volume of the liquid phase. In the first case, the pressure difference across the liquid-gas interface, $\Delta p$, can be imposed through the term 
\begin{equation}
\Psi_\textrm{c} = - \Delta p V_\textrm{l},
\label{eq:free_energy_pressure}
\end{equation} 
where $\Delta p = p_\textrm{l} - p_\textrm{g}$, with $p_\textrm{l}$ and $ p_\textrm{g}$ as the liquid and gas pressures, respectively~\cite{Panter2017}. In this approach, the liquid volume can vary until the system reaches equilibrium in the grand canonical ensemble. In the second case, we can instead constrain the liquid volume through the soft constraint
\begin{equation}
\Psi_\textrm{c} = \frac{1}{2} k \left(V_\textrm{l} - V_0 \right)^2,
\label{eq:free_energy_volume}
\end{equation}
where $k > 0$ is a constant and $V_0$ is the target volume~\cite{Kusumaatmaja2015}. Here, the liquid volume is maintained as approximately the same amount as the target volume, \textit{i.e.}, we are in the canonical ensemble. In either approach, $V_\textrm{l}$ is the actual liquid volume present in the simulation, given by 
\begin{equation}
V_\textrm{l}= \int_V { {\frac{\phi +1}{2}} \textrm{d}V},
\label{eq:volume-liquid}
\end{equation}
where $V$ is the volume of the simulation domain. 

In the simulation, the free energy functional and its derivative are discretized into three types of nodes: bulk fluid nodes, solid nodes and surface nodes (at the solid boundary). The bulk fluid nodes comprise of a cubic lattice with every adjacent node separated by a lattice spacing, $G$, with $G = 0.02$ in simulation units. The interface width for the liquid-gas interface is typically chosen to be $\epsilon = 2G$. The solid and surface nodes are also arranged in a cubic lattice, however, their separation is $G/(2 g_\mathrm{res} - 1)$, where $g_\textrm{res} = 1,2,3,...,n$ denotes the grid resolution. Typically, we use $g_\textrm{res} = 2$. This grid refinement is useful for increasing the accuracy of calculation of the interaction energy density, ${\cal F}_{\textrm{s}}(\textbf{r})$. Every node in the bulk fluid (or solid) is fully occupied with a unit volume of fluid (or solid). At the surface nodes, however, a node is part fluid and part solid, where the corresponding fractions depend on the local surface structure. 

When we compute the numerical integration of interaction energy density ${\cal F}_{\textrm{s}}(\textbf{r})$ as in Eq.~(\ref{eq:lr-external}), all effective liquid-solid interactions are taken into account. These include contributions from bulk fluid-solid, bulk fluid-surface, surface-solid, and surface-surface nodes interactions. The detailed scheme is provided in the Supplementary Material (Sec. S1). In addition, we also employ several periodic images of the solid domain to ensure the long-range liquid-solid interactions are sufficiently accounted. Although the interaction energy density calculation is computationally expensive, particularly for a large domain and a high solid node resolution, it is only calculated once at the start of the simulation. 

Upon the energy minimization routine, the discretized order parameter in the bulk fluid and surface nodes will evolve towards the minimum energy configurations. We employ the L-BFGS algorithm due to its efficiency for problems with a large number of degrees of freedom. For  details on the energy minimization routine, see Refs.~\cite{Kusumaatmaja2015,Panter2017}.

\section{\label{sec.3}Results and Discussion}
\subsection{\label{sec:level2}Wetting on a flat surface for various long-range and short-range liquid-solid interactions}
\subsubsection{\label{sec:level3}Long-range interactions}
\begin{figure*} 
\includegraphics{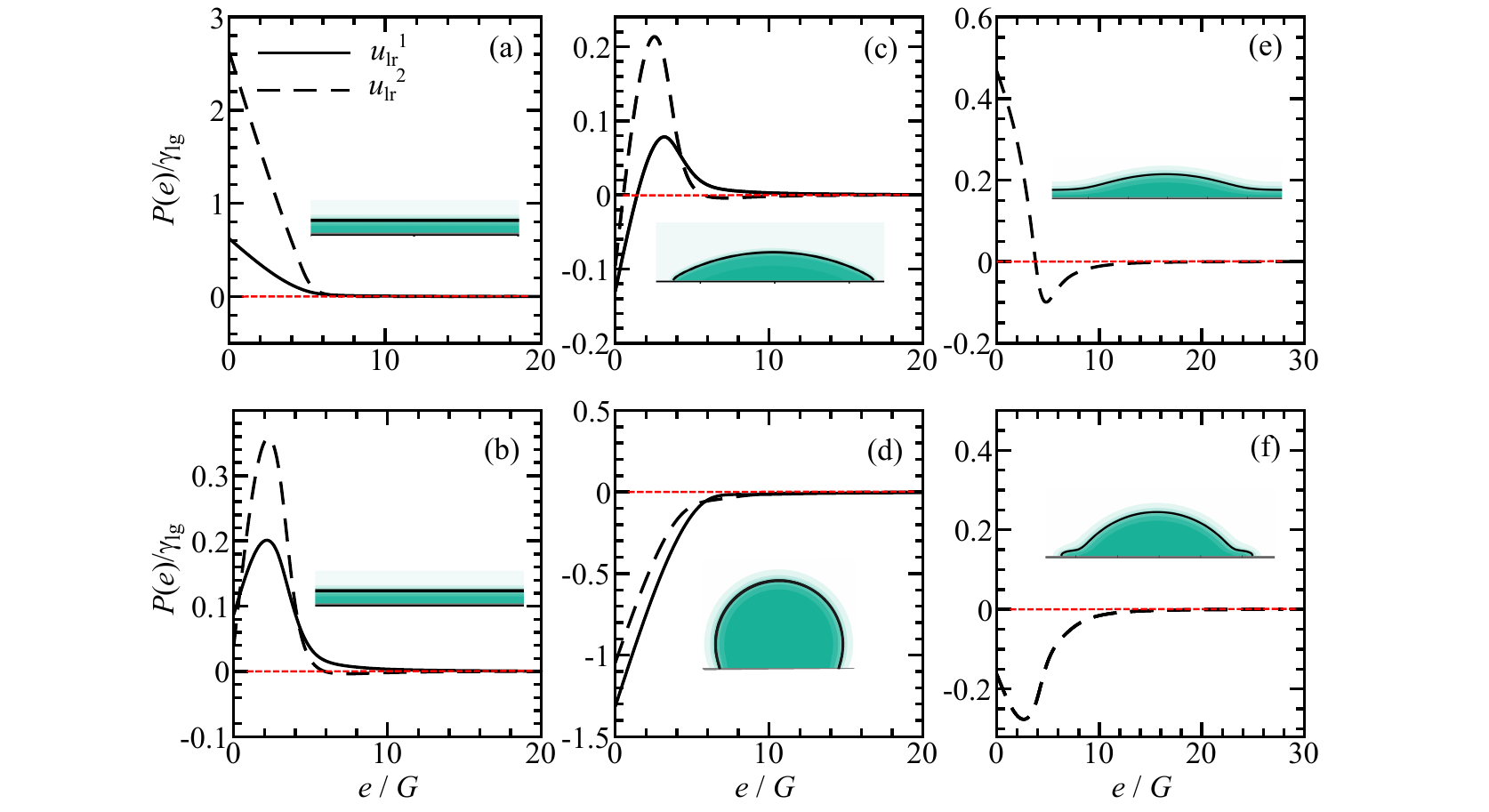}
	\caption{Plots of the reduced effective interface potential $P(e)/\gamma_\textrm{lg}$  of a thin liquid film versus thickness $e$ (in lattice unit \textit{G}) for different wetting states: (a,b) complete wetting, (c,d) partial wetting, and (e,f) pseudo-partial wetting. Insets show the equilibrium state of a droplet on a flat surface. Here, for $u_{\textrm{lr}}^1$ we use (in simulation unit) $\beta = 0.5$ and $\alpha$ of $-8 \times 10^{-4},-4 \times 10^{-4}, -3.2 \times 10^{-4}$ and $1 \times 10^{4}$ for (a-d), respectively. For $u_{\textrm{lr}}^2$, we choose $\sigma = 3$ and $(u_\textrm{o},u_\textrm{p})$ of $(2.4 \times 10^{-6},1 \times 10^{-8})$, $(1.12 \times 10^{-7},1 \times 10^{-8})$, $(1 \times 10^{-7},1 \times 10^{-8})$, $(1.0 \times 10^{-6},1.0 \times 10^{-6})$, $(2 \times 10^{-6},1 \times 10^{-6})$ and $(2 \times 10^{-6},1.4 \times 10^{-6})$ for (a-f), respectively. }
	\label{fig:pe_plot}
\end{figure*} 

To evaluate the effect of the long-range liquid-solid interactions, it is convenient to look at the free energy per unit area of a thin film with a given thickness of $e$, given by~\cite{Brochard-Wyart1991,DeGennes2004} 
\begin{equation}
F\left( e \right) = \gamma_{\textrm{lg}} + \gamma_{\textrm{ls}} + P \left( e \right).
\label{eq:free_energy_theory}
\end{equation}
Here, $P(e)$, called the effective interface potential~\cite{Bonn2009}, is related to the disjoining pressure $\Pi(e)$ in the thin film due to liquid-solid interactions, which vanishes when the thin film is considerably thick $(e \to \infty)$ and acts as the spreading parameter, $S=P(0)$, as the film becomes infinitesimally thin $(e \to 0)$. The disjoining pressure is defined as $ \Pi(e) = -\textrm{d}P(e)/\textrm{d}e $. 

In our model, $\gamma_{\textrm{lg}}=\sqrt{8/9}$, and $\gamma_{\textrm{ls}}$ can be calculated when the film thickness is sufficiently thick such that $\gamma_{\textrm{ls}}=F(e \to \infty)-\gamma_\textrm{lg}$. $P(e)$ can then be determined by evaluating $F(e)$ at varying film thickness. To get the variation of $P(e)$, we simulate a liquid film with small interfacial area on a flat surface to avoid the coexistence between a liquid film and a dry solid. For convenience, here we employ the volume constraint as given in Eq.~(\ref{eq:free_energy_volume}), and vary the film thickness by adjusting the target volume $V_0$ in the simulation. Different variations of $P(e)$ representative of different wetting states are shown in~\figref{pe_plot} (For the disjoining pressure $\Pi(e)$ profiles of these wetting states, see the Supplementary Material, Fig. S2). These capture the profiles previously proposed in the literature, such as by Brochard, \textit{et al.}~\cite{Brochard-Wyart1991}. In the insets, we show an equilibrium state of a sessile droplet placed on a flat surface under the respective wetting states.   

~\figreff[(a,b)]{pe_plot} show the complete wetting case indicated by the positive value of $P(0)$ ($S > 0 $) and the formation of a liquid thin film (insets). In panel (a), we use a large negative $\alpha$ in the effective interaction $u_{\textrm{lr}}^1$. Here, the functional $P(e)$ decreases as the film thickness increases, which results in positive $\Pi(e)$ and $S$. In panel (b), the value of $\alpha$ is less negative than that in panel (a). This makes $P(e)$ non-monotonic leading to negative $\Pi(e)$ at small $e$. The resulting value of $S$ is also smaller but remains positive. A similar profile of $P(e)$ can be seen when using the effective interaction $u_{\textrm{lr}}^2$. Here, $|-u_\textrm{o}| \gg |u_\textrm{p}|$ is used to give a strong attractive interaction near the solid surface to allow the droplet to spread across the surface.

~\figreff[(c,d)]{pe_plot} show the partial wetting case indicated by the negative value of $P(0)$ ($S < 0 $) and droplets with finite contact angles ($\theta < 90^{\circ} $ for panel (c) and $\theta > 90^{\circ} $ for panel (d)), as depicted in the inset of the figures. In panel (c), we still use a negative $\alpha$ in effective interaction $u_{\textrm{lr}}^1$ but the value is smaller than that in the complete wetting case. As a result, the liquid-solid interaction is weaker. Similar to~\figref[(b)]{pe_plot}, $P(e)$ is increasing at small $e$ but decreasing at large $e$. However, the resulting spreading parameter $S$ is negative. If we now switch to positive $\alpha$, $P(e)$ is monotonically increasing with a larger negative $S$, as shown in panel (d). For $\alpha = 0$, $\theta = 90^{\circ}$ and $S < 0$. We can also show the partial wetting case using effective interaction $u_{\textrm{lr}}^2$, as depicted in~\figref[(c,d)]{pe_plot}, where $u_\textrm{p}$ is increased to make a stronger repulsive interaction preventing the droplet from completely spreading.  

If we tune the variable parameters in $u_{\textrm{lr}}^2$ such that the short-range attractive interaction is strong enough to allow the liquid spreading and the long-range repulsive interaction is sufficient to stabilize a droplet, we will obtain a pseudo-partial wetting case, where a droplet is surrounded by a thin liquid film wetting the solid surface, as shown in the insets of~\figref[(e,f)]{pe_plot}. The spreading parameter $S$ can be negative or positive~\cite{Brochard-Wyart1991,Yeh19995}, and the $P(e)$ profile is characterized by a minimum at a certain $e$. In panel (e), the attractive term is quite strong at short ranges (due to large $|-u_\textrm{o}|$ ) that $P(e)$ is decreasing. Since $S > 0$, the thin film extends indefinitely. At long ranges, the repulsive term becomes more dominant (due to moderate value of $|u_\textrm{p}|$) to change the direction of $P(e)$ to be increasing. The droplet formed in this condition has a lower contact angle, as seen in the inset. When the strength of repulsive terms is increased but the attractive term is kept unchanged, the decreasing trend of $P(e)$ at short ranges reduces and the increasing trend of $P(e)$ at long ranges increases, as shown in panel (f). As can be seen, the contact angle of the droplet is larger (see inset). Moreover, since $S < 0$, the thin film does not extend indefinitely, as illustrated in the inset. The pseudo-partial wetting case cannot be obtained with effective interaction $u_{\textrm{lr}}^1$. 

\subsubsection{\label{sec:level3}Short-range interactions}
\begin{figure}
\centering
\includegraphics{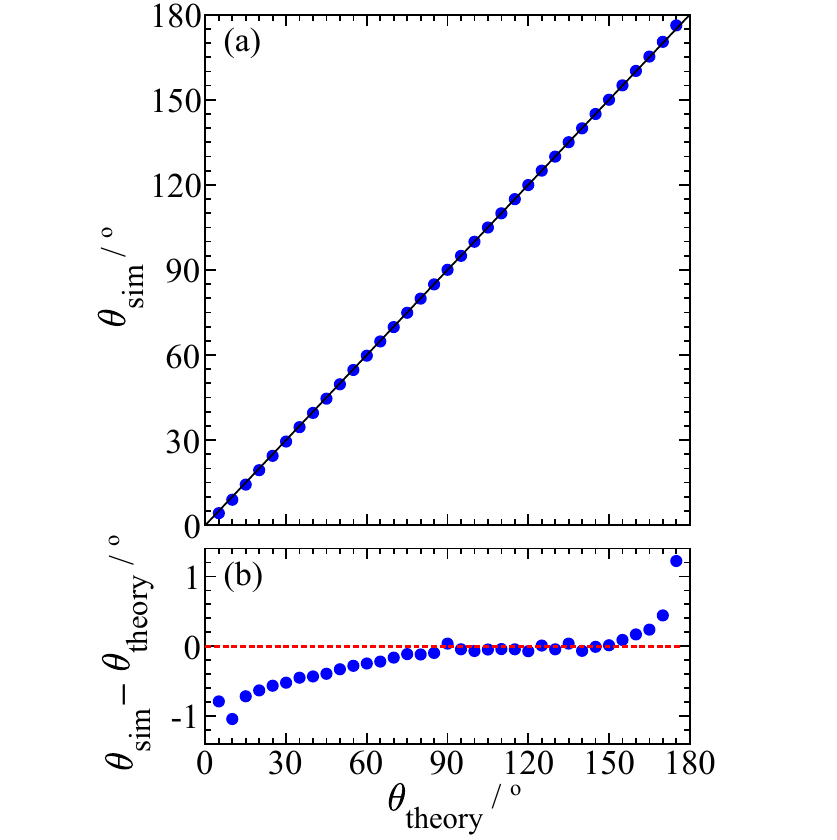}
\caption{(a) Comparison of the input ($\theta_{\textrm{theory}}$) and the measured ($\theta_{\textrm{sim}}$) contact angles when the short-range interaction is used in the surface energy density. (b) Absolute error of the measured contact angle. Excellent agreement is obtained with an error of $ < 1 ^{\circ}$.}
\label{fig:ca_short_range}
\end{figure}
When considering large-scale wetting phenomena, the long-range liquid-solid interactions discussed in the preceding sub-section are often not directly relevant, as they occur at much smaller length scales. The short-range surface energy density implemented in the free energy is directly related to contact angle at the surface via Eq.~(\ref{eq:free_energy_surface_SR_2}). Similar short-range energy densities have previously been demonstrated in the literature~\cite{Panter2019}. Here, the main difference is the quartic form of $f(\phi_\textrm{s})$. In~\figref[(a)]{ca_short_range}, we compare the measured contact angle of a sessile 2D drop from the simulation, labelled $\theta_{\textrm{sim}}$, with the input contact angle, $\theta_{\textrm{theory}}$. To measure the contact angle, we fit a circular arc to the drop profile. We found an excellent accuracy of the contact angle with the error of $< 1 ^{\circ}$ (\figref[(b)]{ca_short_range}). Such accuracy is superior compared to a range of frequently used forms of $f(\phi_\textrm{s})$ \cite{Huang2015} including linear and cubic models. The comparisons between the different forms of $f(\phi_\textrm{s})$ are provided in the Supplementary Material, Table S1 and Fig. S3.

\subsection{\label{sec:level2}Wetting on grooved surfaces}

Our next investigation is the wetting behavior of liquid on a structured surface. In this context, we consider a long periodic grooved surface with groove width of $d$, depth $h$, and wall barrier width $w$, as shown in~\figref[(a)]{system_groove}. To reduce the simulation cost, it is only necessary to simulate a single groove unit cell with periodic boundary condition being applied in the \textit{x} and \textit{y} directions to capture the periodicity of the grooves. In this work, the simulation domain size is chosen to be $N_x = w+d$, $N_y = 6G$ and $N_z = h+50G$. The groove dimension is taken as $w = h = 50G$ and $d=20G$, unless stated otherwise.  The typical liquid-solid interaction energy densities due to long-range interactions across the system are shown in~\figref[(b)]{system_groove} for the effective interactions $u_{\textrm{lr}}^1$ and $u_{\textrm{lr}}^2$. Here, the energy density is scaled by $\gamma_{\textrm{lg}}/G$. For $u_{\textrm{lr}}^1$, the parameter $\alpha$ is taken as a negative value, hence liquid and solid experience an attractive interaction which is higher at the surface and decays towards zero farther from the surface. This is depicted in~\figref[(c)]{system_groove} for complete and partial wetting cases at $x = 0$ and varying $z$. The decay rate of $u_{\textrm{lr}}^1$ depends on the parameter $\beta$. The higher $\beta$, the slower the decay and the longer the interaction tail. For $u_{\textrm{lr}}^2$, the attractive interaction only occurs near the surface, and the interaction becomes repulsive in the bulk of the liquid, as depicted in~\figref[(c)]{system_groove} for the pseudo-partial wetting case. It is also worth noting that for both effective interactions, the liquid-solid interaction is stronger at the bottom corners and weaker at the top corners of the barrier wall, consistent with observations from MD~\cite{Liu2022} and DFT~\cite{Giacomello2016, Singh2022} simulations.
\begin{figure}
	\centering
	\includegraphics{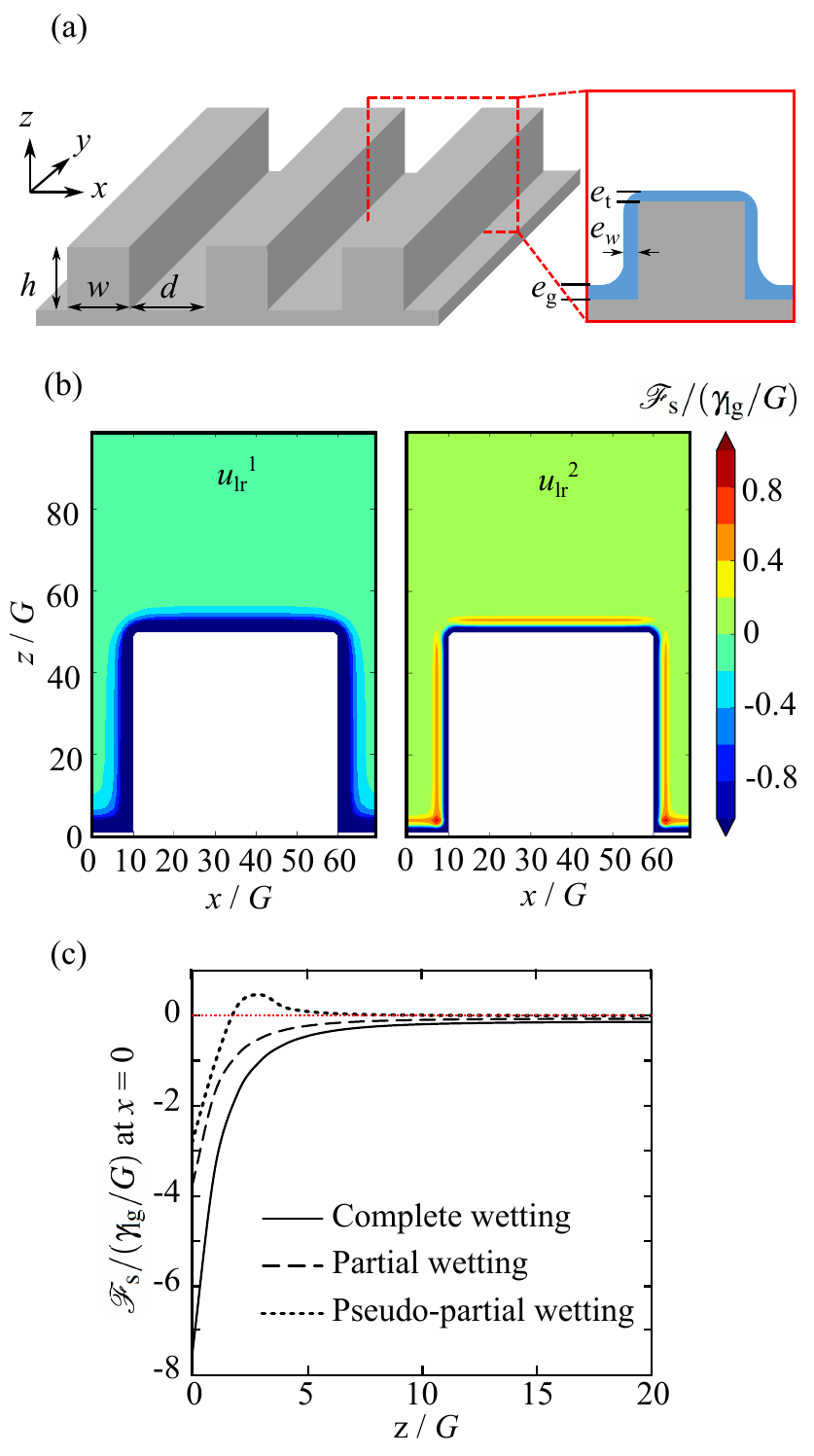}
	\caption{(a) (left) 3D sketch of the groove patterned surface and (right) a 2D slice of a unit cell of the surface. (b) Contour plots of the typical profile of reduced interaction energy density ${\cal F}_{\textrm{s}}/(\gamma_{\textrm{lg}}/G)$ due to effective interactions $u_{\textrm{lr}}^1$ (left) and  $u_{\textrm{lr}}^2$ (right) for a 2D slice of the system. Color bars show the value of the energy density. (c) Reduced interaction energy density ${\cal F}_{\textrm{s}}/(\gamma_{\textrm{lg}}/G)$ plotted as a function of position in the \textit{z} direction taken at $x = 0$ for the complete and partial wetting cases using $u_{\textrm{lr}}^1$ and the pseudo-partial wetting case using $u_{\textrm{lr}}^2$.}
	\label{fig:system_groove}
\end{figure}
\begin{figure*}
	\centering
	\includegraphics{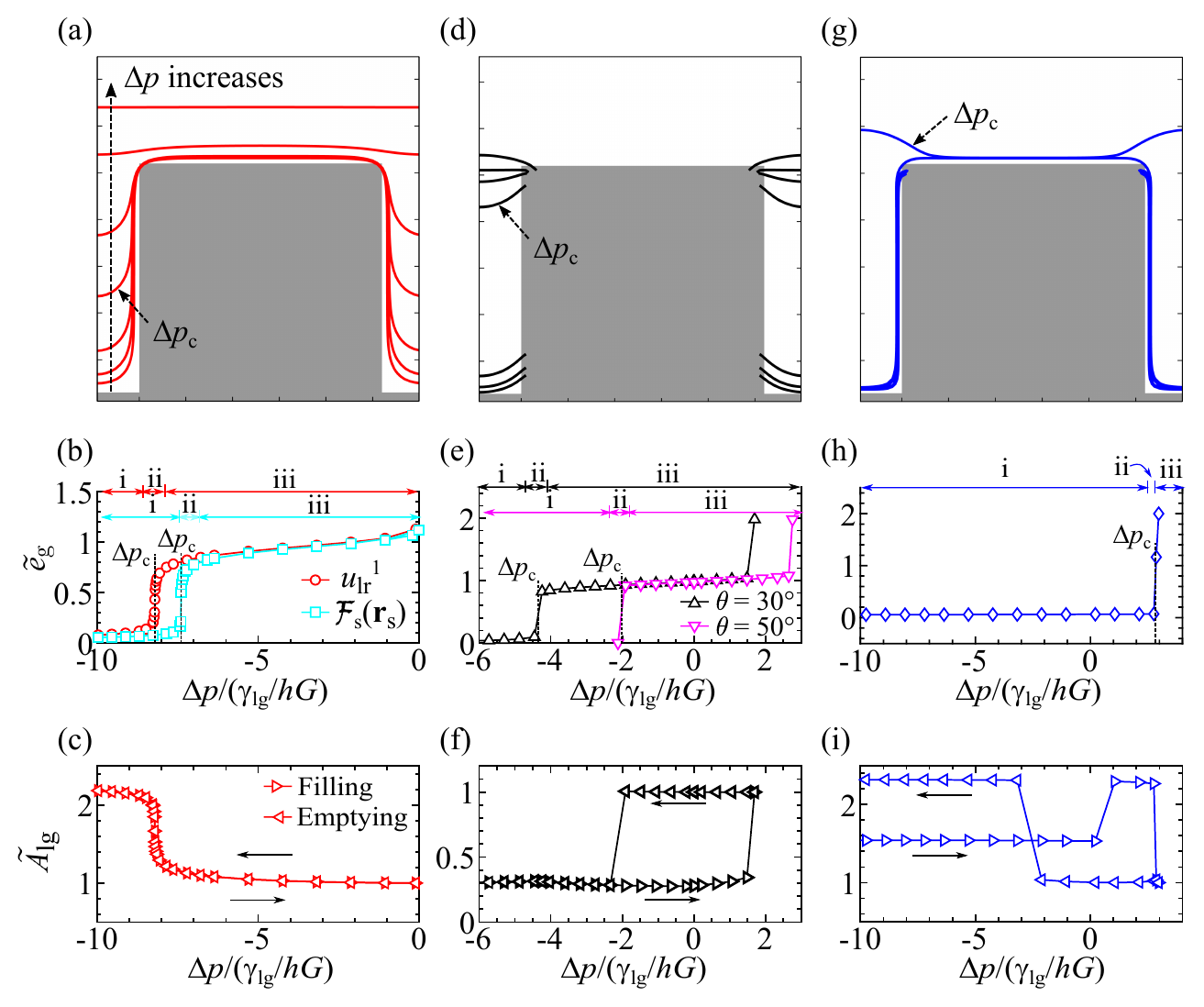}
	\caption{Liquid filling processes on the grooved surface for complete (a, b, c), partial (d, e, f) and pseudo-partial wetting (g, h, i) cases. Here, $d/h = 0.4$. Each contour line in panels (a, d, g) corresponds to the liquid-gas interface ($\phi = 0$) at an increasing $\Delta p$. Panels (b, e, h) show the plot of reduced liquid film thickness measured at the middle of the gap $\widetilde{e}_\textrm{g}$ against reduced pressure difference $\Delta p/(\gamma_{\textrm{lg}}/hG)$. In panel (b) we show the complete wetting case when using ${\cal F}_\textrm{s}(\textbf{r})$ with $u_\textrm{lr}^1$ and  ${\cal F}_\textrm{s}(\textbf{r}_\textrm{s})$. In panel (e) we show the partial wetting case for two contact angles, $\theta = 30^{\circ}$ and $\theta = 50^{\circ}$. (i-iii) denote the liquid filling stages with increasing $\Delta p$: (i) pre-filling, (ii) capillary filling and (iii) post-filling. The filling transition occuring at $\Delta p_\textrm{c}$ is indicated in the figures. Panels (c, f, i) are the plots of reduced liquid-gas interface area $\widetilde{A}_\textrm{lg}$ against $\Delta p/(\gamma_{\textrm{lg}}/hG)$. Right and left arrows indicate the liquid filling path and liquid emptying path for increasing and decreasing $\Delta p/(\gamma_{\textrm{lg}}/hG)$, respectively. The solid line is to guide the eyes.} 
	\label{fig:filling_corrugated}
\end{figure*}

\subsubsection{\label{subsec:level1}Complete wetting}

~\figreff[(a,b,c)]{filling_corrugated} shows the filling and emptying transition as the pressure $\Delta p$ is varied, for the case of complete wetting. Here, we use $u_{\textrm{lr}}^1$ for the long-range interaction with $\alpha = -8 \times 10^{-4}$ and $\beta = 0.5$. Similar results are obtained when $u_{\textrm{lr}}^2$ is used. Upon increasing the liquid pressure, liquid begins to fill the grooved surface, as shown in~\figref[(a)]{filling_corrugated}. We can categorize the liquid filling process into three stages~\cite{Hofmann2010}, namely (i) pre-filling, (ii) capillary filling, and (iii) post-filling, which occur after one and the other with increasing liquid pressure. 

The pre-filling stage occurs at large negative $\Delta p$, which means the pressure in the liquid is much lower than that in the gas phase. Here, the liquid forms a thin film that follows the shape of the groove structure. The thickness of the film depends on the strength of the interaction (parameter $\alpha$ in $u_{\textrm{lr}}^1$) and increases as the liquid pressure is increased. The dependency is well approximated by $e \propto (2\alpha/\Delta p)^{-1/3}$ at the bottom, top and sides of the barrier wall, where $\alpha$ can be associated with the Hamaker constant ~\cite{Lipowsky1985, Laska2021}. Assuming the grooves have the dimension of order of hundreds nm, the values of $\alpha$ in our simulation translate $10^{-20}$ to $10^{-19}$ J, which are the typical values of the Hamaker constant~\cite{israelachvili2011intermolecular}. 

The key to understanding the capillary filling stage lies in changes to the film thickness in the bottom corners of the groove. As the menisci in the corners grow and approach $d/2$ in size, liquid from either side merges and rapidly fills the gap. The liquid interface then rises up from the bottom of the groove. This can be seen from the sudden increase in liquid film thickness as calculated at the middle gap, $\widetilde{e}_\textrm{g}$ (\figref[(b)]{filling_corrugated}) and from the sudden decrease of the liquid-gas interfacial area, $\widetilde{A}_\textrm{g}$ (\figref[(c)]{filling_corrugated}). We define the critical pressure $\Delta p_\textrm{c}$ as the pressure value with the largest gradient in the $\widetilde{e}_\textrm{g}$ and $\widetilde{A}_\textrm{g}$ plots. At this capillary filling stage, the growth of thin film at the side walls and top of the barrier wall still follows $e \propto (1/\Delta p)^{-1/3}$.

The liquid, however, does not immediately fill the whole gap of the groove. In the post-filling stage, with increasing pressure, the liquid-gas interface between the barrier walls starts to smooth out until it becomes flat. The thin film thickness at the top of the barrier wall also increases more rapidly compared to the pre-filling and capillary filling stages. This occurs when $\Delta p$ has small negative values, which means $p_\textrm{l} \to p_\textrm{g}$. As $p_\textrm{l} > p_\textrm{g}$, the film thickness increases to infinity as liquid fills up the whole domain. 

If $\Delta p$ is reversed from positive to large negative values, the liquid will be emptied from the grooved surface. Upon decreasing $\Delta p$, the liquid-gas interface follows the reverse path as the liquid fills the groove surface (\figref[(c)]{filling_corrugated}). Therefore, the liquid filling does not exhibit hysteresis behavior for the complete wetting case. Recently, filling transitions have been investigated via DFT~\cite{Singh2022}. It was also observed that the filling transition is mediated by the growth of the menisci in the bottom corners of the groove. However here, we are also able to show the contribution of the films on the sidewalls and top of the barrier wall.

Next, we want to compare the effect of long-range and short-range liquid-solid interactions on the liquid filling transition. In this case, we use the effective interaction $u_{\textrm{lr}}^1$ in the interaction energy density ${\cal F}_{\textrm{s}}(\textbf{r})$ for the former and ${\cal F}_{\textrm{s}}(\textbf{r}_\textrm{s})$ as in Eq.~(\ref{eq:free_energy_surface_SR_1}) for the latter. Although the filling behavior for both interactions is qualitatively the same, the filling transition occurs at different critical pressures (\figref[(b)]{filling_corrugated}). This is because the liquid film thickness at the wall is different, which changes the effective separation between the walls. The critical pressure dependency can be inferred from the Kelvin equation, in which $\Delta p_c$ is expected to be inversely proportional to the effective wall separation. 

The liquid film formed due to long-range interactions is thicker than that due to short-range interactions. For the former, the contribution of the liquid-solid interaction is determined by how long the tail of the decaying interaction is until it becomes essentially zero. This is controlled by the parameter $\beta$. The higher $\beta$, the longer the tail. As a result, the interaction with higher $\beta$ forms a thicker liquid film at the wall. For the short-range interaction, the liquid-solid interaction is assumed to occur only at the surface of the solid. Therefore, there is no liquid-solid interaction contribution farther from the surface. Hence, the liquid film is thinner, and the filling transition occurs for larger $\Delta p$. 

\subsubsection{\label{subsec:level2}Partial wetting}

We now turn our attention to the partial wetting case ($\theta > 0^{\circ}$). The results presented here employ the short-range interaction. Equivalent results are obtained for the long-range interactions once the contact angles are mapped. The liquid filling behavior for the partial wetting case is illustrated in~\figref[(d,e,f)]{filling_corrugated}. In the same manner as the complete wetting case, we can also group the filling process into (i) pre-filling, (ii) capillary filling and (iii) post-filling stages. The capillary filling stage is also marked by a critical pressure at $\Delta p_\textrm{c}$. Here, we have to divide our discussion into two scenarios~\cite{Malijevsky2018}, which are for $\theta < 45^{\circ} $ and for $\theta > 45^{\circ}$ . 

For $\theta < 45^{\circ} $, in the pre-filling stage, liquid condensation could be nucleated at the corner of the groove forming menisci at a large negative $\Delta p$ (\figref[(d)]{filling_corrugated}). In this case, the corner menisci grow as the liquid pressure increases until they merge as a single meniscus. Once this has happened, we enter the capillary filling stage, in which the liquid starts filling the gap while maintaining the shape of the meniscus. In sharp interface models, at a certain $\Delta p_\textrm{c}$, the filling transition occurs as signified by an abrupt increase in $\widetilde{e}_\textrm{g}$. $\Delta p_\textrm{c}$ can be predicted using the Kelvin equation, as will be discussed in Section~\ref{sec:level4}. Using the diffuse interface model in the present study results in a rounding of this first-order phase transition. However, as is shown in the Supplementary Material (Sec. S4), this effect is marginal if there is a suitable separation of length scales (at least a factor of 10) between the diffuse interface width and the wall height. Thus, as is shown in~\figref[(e)]{filling_corrugated}, the partial wetting filling transition is still sharp compared to the complete wetting case. After the filling transition occurs, the liquid again does not completely fill the gap, as in the complete wetting case, but both ends of the liquid-gas interface are pinned in the top edge of the wall. In the post-filling stage, the meniscus starts to flatten as $p_\textrm{l} \to p_\textrm{g}$. When $\Delta p$ turns positive, the curvature of the meniscus also turns sign from negative to positive. As the liquid manages to overcome the contact line pinning, it fills up all the gas phase. 

For $\theta > 45^{\circ}$, the pre-filling stage is marked by a gas-like phase with a completely dry solid (\figref[(e)]{filling_corrugated}). The corner menisci do not form, and the filling transition immediately occurs when the pressure has reached $\Delta p_\textrm{c}$. The liquid will then be pinned at the top edge of the groove with smaller curvature due to higher $\theta$. The post-filling stage is then similar to that for $\theta < 45^{\circ}$ except that the positive curvature at positive $\Delta p$ could grow larger in size before it overcomes the contact line pinning and fills all of the gas phase. This means that $\Delta p$ at which the liquid fills the gas phase occurs at a larger value than that for $\theta < 45^{\circ}$. 

~\figreff[(f)]{filling_corrugated} shows liquid filling and emptying paths for increasing and decreasing $\Delta p$. The hysteresis behavior is clearly pronounced. Due to contact line pinning at the top corner of the walls, during the filling process, the meniscus curvature changes from negative to positive as $\Delta p$ increases. During the emptying process, however, upon decreasing  $\Delta p$ the liquid continues to wet all the surface and maintains a flat liquid-gas interface until a significantly lower pressure difference. Once the top of the wall is fully dewetted, the liquid gets pinned at the top corners with negative curvature. The liquid emptying path then follows along the same path as the liquid filling (See the Supplementary Material, Fig. S5, for the snapshots of configurations during the filling and emptying process). This hysteresis behavior in partial wetting case has also been reported elsewhere~\cite{Malijevsky2012,Rascon2013}. 

\subsubsection{\label{subsubsec:level3}Pseudo-partial wetting}

~\figreff[(g,h,i)]{filling_corrugated} shows the liquid filling behavior on a grooved surface for the pseudo-partial wetting case. The effective interaction $u_{\textrm{lr}}^2$ is used in the interaction energy density ${\cal F}_{\textrm{s}}(\textbf{r})$. The magnitude of parameter $-u_\textrm{o}$ controls the strength of the attractive interaction. To obtain a pseudo-partial wetting state, a large enough $-u_\textrm{o}$ is employed to get a liquid film near the surface. The parameter $\sigma$ controls the thickness of the liquid film. Here, we use $\sigma = 3$ and $u_\textrm{o}= -5 \times 10^{-7}$ and $u_\textrm{p} = 2 \times 10^{-7} $.

In the pre-filling stage (at large negative $\Delta p$), in contrast to the partial wetting case at the same $\Delta p$, the liquid wets the bottom surface of the groove and the side walls forming a liquid film, but leaves the top of the barrier wall dry as the liquid film is pinned at the top edges of the wall. At the bottom corners of the groove, the meniscus of liquid condensation is not as pronounced as it is for the full and partial wetting case. This is because the interaction at the surface near the corner slightly reduces due to the effect of the repulsive term in the effective interaction $u_{\textrm{lr}}^2$. As the liquid pressure increases the liquid overcomes the contact line pinning at the top edges and covers the top of the wall. This is shown in~\figref[(i)]{filling_corrugated} (right-pointing triangle), in which the liquid-gas interface area increases abruptly at $\Delta p/(\gamma_{\textrm{lg}}/hG) \approx 0.3$. The bottom corner menisci only slightly grow with increasing pressure, unlike for the complete and partial wetting cases where they grow and merge as their size approaches $d/2$.

In the capillary filling stage, the critical pressure for the filling transition occurs sharply at positive $\Delta p_\textrm{c}$ (indicated in~\figref[(h)]{filling_corrugated}). The sharp transition applies for narrow and wide groove widths. Compared to the negative $\Delta p_\textrm{c}$ observed for complete and partial wetting cases, this suggests the filling transition is more energetically expensive for the pseudo-partial wetting case. At the critical pressure $\Delta p_\textrm{c}$, liquid fills the gap, and it forms a droplet in the middle of the gap coexisting with the liquid film on top of the barrier wall (\figref[(g)]{filling_corrugated}). Such coexistence is reminiscent of the morphology observed on a flat surface.  However, the range of stability of the droplet is limited. With increasing pressure in the post-filling stage (with positive $ \Delta p $), the droplet becomes unstable and the liquid fills the simulation domain. 

The hysteresis behavior is also clearly observed in the pseudo-partial wetting case, as shown in~\figref[(i)]{filling_corrugated} (See the Supplementary Material, Fig. S5, for snapshots of configurations during the filling and emptying processes). During the filling process, the contact line pinning at the top corners of the walls allows the liquid to form a droplet bulge in the middle of the gap as $\Delta p$ increases. Upon decrasing $\Delta p$, however, the droplet bulge slowly flattens until the liquid filling the gap abruptly drains, leaving a liquid film that follows the shape of the groove structure. The top of the wall remains covered by a liquid film, hence we find higher $\widetilde{A}_\textrm{lg}$ than in the liquid filling path.

\subsubsection{\label{sec:level4}Critical pressure scaling with groove width} 

\begin{figure}[ht]
\includegraphics{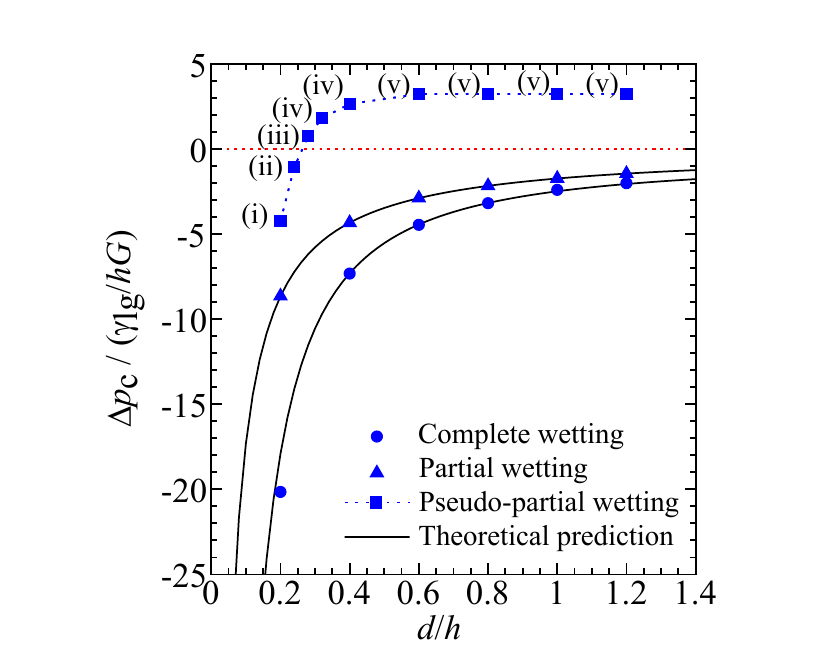}
\caption{Reduced critical pressure $\Delta p_\textrm{c}/(\gamma_{\textrm{lg}}/hG)$ as a function of separation-to-post height ratio $d/h$ for complete, partial, and pseudo-partial wetting cases. The error bars for the data points are not visible as they are comparable to the marker size. The solid curve is the theoretical prediction for partial and complete wetting cases given by Eq.~(\ref{eq:model_partial}) with $\theta = 30^\circ$ and Eq.~(\ref{eq:model_full_corrected}), respectively. (i-v) denote specific $d/h$ at which different scenarios of the filling transition occur for the pseudo-partial wetting case (see text and~\figref{separation_effect_pseudo}).}
\label{fig:separation_effect}
\end{figure}

In this section, we will now consider how the critical pressure for the filling transition depends on the groove width, $d$. We will begin by considering the partial wetting case. To describe the critical pressure quantitatively, we can use the following argument. During the transition, the groove will experience a change of liquid volume $\Delta V$, accompanied by a change of liquid height in the groove by $h$. As such, the change in the total free energy is given by
\begin{equation}
\Delta E=-\Delta p \Delta V + \gamma_\textrm{lg}\Delta A_\textrm{lg}+\gamma_\textrm{ls}\Delta A_\textrm{ls}+\gamma_\textrm{gs}\Delta A_\textrm{gs},
\label{eq:model-1}
\end{equation}
where $\Delta A_\textrm{lg}$, $\Delta A_\textrm{ls}$ and $\Delta A_\textrm{gs}$ are the changes in liquid-gas, liquid-solid and gas-solid interface areas, respectively. During the filling transition, the liquid-gas interface remains nearly constant, hence $\Delta A_\textrm{lg}\approx 0$. $\Delta V$, $\Delta A_\textrm{ls}$ and $\Delta A_\textrm{gs}$ can be approximated by $ d^2h $, $2dh$ and $-2dh$, respectively.  Using Young's equation, we can rearrange Eq.~(\ref{eq:model-1}) to obtain 
\begin{equation}
\Delta E=-\Delta p d^2h - 2\gamma_\textrm{lg} \cos \theta dh.
\label{eq:model-2}
\end{equation}
The critical pressure corresponds to the case where $\Delta E = 0$, leading to a relation
\begin{equation}
  \Delta p_\textrm{c} = -\frac{2\gamma_\textrm{lg} \cos \theta}{d}.  
\label{eq:model_partial}
\end{equation}
This equation has the same form as the Kelvin equation and, as shown in~\figref{separation_effect}, it captures the critical pressure obtained in the simulation accurately. 

A similar argument can be applied for the complete wetting case with $\theta = 0$. However, to account for the effect of the liquid film at the wall, a correction term of $3e_\textrm{w}$, where $e_\textrm{w}$ is the film thickness at the wall, needs to be added because the effective wall separation is not equal to $d$, as proposed by Derjaguin~\cite{derjaguin1992theory}. Therefore, the critical pressure becomes~\cite{Evans1986}
\begin{equation}
  \Delta p_\textrm{c} = -\frac{2\gamma_\textrm{lg}}{d-3e_\textrm{w}},   
  \label{eq:model_full_corrected}
\end{equation}
where $e_\textrm{w}$ depends on $\Delta p$ through the relation $e_\textrm{w} \propto (1/\Delta p)^{-1/3}$. The comparison of the critical pressure between simulation and theoretical predictions for complete wetting case is also shown in~\figref{separation_effect}. It shows a good agreement to a very narrow gap although there is a slight deviation for $d/h = 0.2$ because the interface width of our diffuse liquid-gas interface becomes comparable to $d$. 

\begin{figure}[h]
\includegraphics{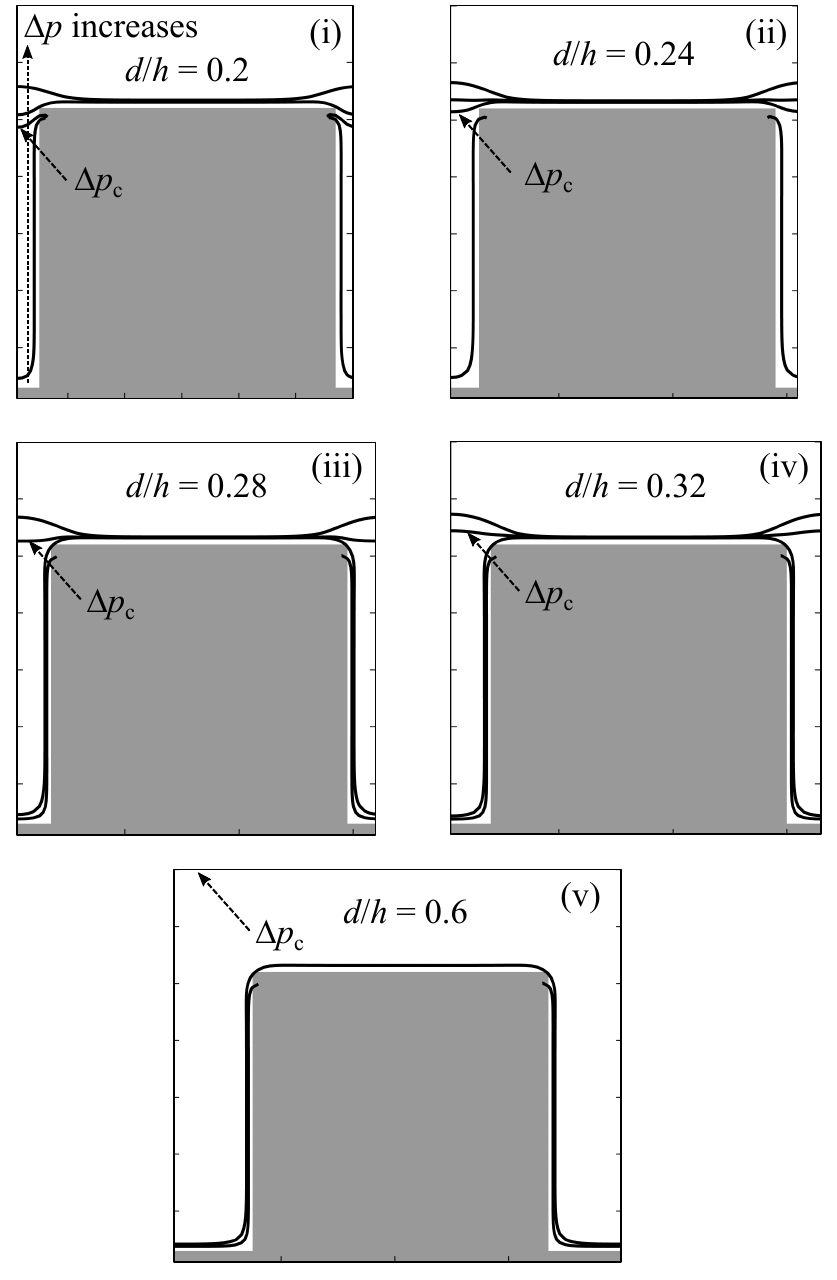}
\caption{Liquid filling states at different groove widths for the pseudo-partial wetting case. Panels (i-v) correspond to different scenarios as indicated in~\figref{separation_effect} and explained in the main text. In panel (v), the liquid fills up the whole simulation domain when the liquid filling transition occurs; hence the liquid-gas interface at $\Delta p_\textrm{c}$ lies at the top of the domain.}
\label{fig:separation_effect_pseudo}
\end{figure}

The pseudo-partial wetting case, however, cannot be captured by a relation akin to Eq.~(\ref{eq:model_partial}) or Eq.~(\ref{eq:model_full_corrected}). We argue that this is because the corner menisci are not so apparent during the capillary filling stage and do not merge into a single meniscus before the filling transition occurs. Therefore, $d$ does not affect $\Delta p_\textrm{c}$. This can be observed for $d/h > 0.6$ in~\figref{separation_effect}. In this scenario, the liquid fills up the simulation when the filling transition occurs, as illustrated in ~\figref[(v)]{separation_effect_pseudo}.
When $d$ is very small ($d/h < 0.6$), however, $\Delta p_\textrm{c}$ starts to be dependent on $d$, but it still does not obey Eq.~(\ref{eq:model_partial}). Instead, we find this variation is accompanied by non-trivial changes in the morphological pathway during the filling transition. With increasing groove width, five distinct pathways are identified, illustrated in~\figref{separation_effect_pseudo}, and indicated in~\figref{separation_effect}: (i) liquid fills the gap forming a liquid-gas interface with a negative curvature while keeping the top of the wall dry, (ii) the same as scenario (i) except that top of the wall is covered by a liquid film, (iii) the same as scenario (ii) but the liquid film at the top of the wall is formed before the capillary filling stage, (iv) the same as scenario (iii) but the liquid-gas interface curvature is positive, and (v) liquid fills the system at the critical pressure.

\section{\label{sec.4}Conclusion}
In summary, we have presented systematic numerical studies of liquid filling on grooved surfaces using a phase field method. We consider both short-range and long-range liquid-solid interactions. The latter include purely repulsive and attractive interactions, and more complex interactions with short-range attraction and long-range repulsion. To the best of our knowledge, such versatility allows us to capture complex disjoining pressure profiles for the first time in a phase field approach, in agreement with previous works using atomistic modelling~\cite{Sethi2022,Malijevsky2013} and analytical theory~\cite{Brochard-Wyart1991,Yeh19995}, which in turn give rise to complete, partial, and pseudo-partial wetting states. In this work, we have also introduced a quartic polynomial to switch the interaction energy density between the liquid and gas phases ($f(\phi)$ in Eq.~(\ref{eq:lr-quartic})). This polynomial prevents enrichment of the liquid and/or gas phases on the solid surface, and it leads to more accurate contact angle calculations compared to the linear and cubic forms previously used in the literature. 

We rationalize the liquid filling process on grooved surfaces into three stages: (i) pre-filling, corresponding to the growth of the thin film around the structure (for complete or pseudo-partial wetting), or liquid menisci in the bottom corners of the groove (for partial wetting); (ii) capillary filling, where there is a rapid increase of liquid volume in the groove marked by a critical pressure; and (iii) post-filling, typically signified by the flattening of the liquid-gas interface before liquid completely fills the whole domain. Comparing the results for complete, partial, and pseudo-partial wetting, we find there is no contact line pinning for the complete wetting case and the liquid filling and emptying trajectories are reversible. In contrast, we observe clear hysteretic behaviour for partial and pseudo-partial wetting, caused by the coexistence of two metastable states over a pressure range. In the partial wetting case, late in the filling transition, pinning of the interface on the top corner of the wall leads to a state that remains metastable over a range of positive pressures. Coexisting with this is the unpinned state, which at positive pressures sees liquid completely fill the system. In the pseudo-partial wetting case, the origin of metastability is different. Here, repulsive interactions in the centre of the groove energetically penalise partial filling of the groove. Instead, either the groove remains almost empty, or the groove is full. Considering the critical pressure, although the diffuse interface marginally rounds the first-order transitions, both the complete and partial wetting cases follow a Kelvin-like equation for its dependence on the groove width: large and negative at small widths, and plateaus to zero for large widths. The pseudo-partial wetting case, however, is different. At large widths, the critical pressure is positive and constant. We find the critical pressure does depend on groove width for smaller widths, and interestingly, this is accompanied by morphological changes in the trajectories of the liquid filling process. 

There are a number of exciting avenues for future work. Here, we have considered grooved surfaces. It is straightforward to extend the study to other, more complex surface geometries such as re-entrant geometries, seesaws, hierarchical posts, or even non-symmetric structures which have extensively been harnessed in wetting applications. Another possible direction is to consider the liquid dynamics, beyond the quasi-static results presented here. There are some limitations, however, of the phase field method used in this study. The present method does not include the interfacial fluctuation effects, which have been shown to occur at nanoscale~\cite{Aarts2004} and captured by atomistic simulations~\cite{Parry2001}. To capture these phenomena, one possible route is to couple the phase field model here with fluctuating hydrodynamics methods~\cite{Chaudhri2014,Gallo2018}. Finally, we hope the simulation results will inspire experimental studies to verify our theoretical predictions.

\section*{\label{sup.mat}Supplementary material}
The supplementary material contains (i) our scheme for the calculation of ${\cal F}_{\textrm{s}}(\textbf{r})$; (ii) the corresponding disjoining pressure profiles for the effective interfacial potential shown in~\figref{pe_plot}; (iii) comparison of contact angle results for different forms of $f(\phi_\textrm{s})$; (iv) a short discussion on the rounding of first-order phase transition due to the diffuse interface; and (v) liquid-gas interface configurations during the filling and emptying processes for partial and pseudo-partial wetting.

\begin{acknowledgments}

F.O. acknowledges a BPPLN scholarship from the Directorate General for Science Technology and Higher Education, Republic of Indonesia. H.K. and J.R.P. would like to thank EPSRC (EP/V034154/1) for funding.
\end{acknowledgments}

\section*{Author Declarations}
\subsection*{Conflict of Interest}
The authors have no conflict to disclose.
\subsection*{Author Contributions}
\textbf{Fandi Oktasendra:} Conceptualization (equal); Investigation (lead); Methodology (equal); Writing - Original Draft (equal). \textbf{Arben Jusufi:} Conceptualization (equal); Funding acquisition (lead); Writing - Review \& Editing (equal). \textbf{Andrew R. Konicek:} Conceptualization (equal); Funding acquisition (equal); Writing - Review \& Editing (equal). \textbf{Mohsen S. Yeganeh:} Conceptualization (equal); Funding acquisition (equal); Writing - Review \& Editing (equal). \textbf{Jack R. Panter:} Conceptualization (equal); Methodology (equal); Supervision (equal); Writing - Original Draft (equal). \textbf{Halim Kusumaatmaja:} Conceptualization (equal); Funding acquisition (lead); Supervision (lead); Writing - Original Draft (equal).

\section*{Data Availability Statement}
Data supporting the findings in this study are available from the corresponding authors upon reasonable request. 


\nocite{*}
\bibliography{aipsamp}

\end{document}